# LLMs and Childhood Safety: Identifying Risks and Proposing a Protection Framework for Safe Child-LLM Interaction

**Junfeng Jiao**
Urban Information Lab
University of Texas at Austin
jjiao@austin.utexas.edu

**Saleh Afroogh** [1]
Urban Information Lab
University of Texas at Austin
saleh.afroogh@utexas.edu

**Kevin Chen**
Urban Information Lab
University of Texas at Austin
xc4646@utexas.edu

**Abhejay Murali**
Urban Information Lab
University of Texas at Austin
abhejay.murali@utexas.edu

**David Atkinson**
Allen Institute for AI (AI2)
Seattle, USA
davida@allenai.org

**Amit Dhurandhar**
IBM Research
Yorktown Heights,USA
adhuran@us.ibm.com

## Abstract

This study examines the growing use of Large Language Models (LLMs) in child-centered applications. With the emergence of widespread usage for artificial intelligence (AI), comes safety and ethical concerns surrounding AI model prompt outputs in areas of bias, inappropriate content, and cultural insensitivity relevant to children. The absence of standardized ethical structures to reduce these risks is a challenge and we intend to tackle through a systematic literature review, by pinpointing empirical issues such as toxicity and ethical violations in AI results. In addition, to enhance efforts in tackling these problems, this paper suggests a safety framework for secure Child-LLM interactions, employing metrics for content safety and ethical behavior. The goal of our framework is to gauge an understanding on the status of safe AI usage for children and aim to provide practical tools for evaluating LLM safety, offering guidance for developers, policymakers, and educators to incorporate responsible AI usage for children and further endeavors for safety.

[1] Corresponding author

# I. Introduction

Children's learning and interaction with digital technologies are being changed by LLMs included into educational environments. These sophisticated artificial intelligence systems are being applied more and more as instruments for improving tailored support, creativity, and education. Children's particular cognitive, emotional, and developmental weaknesses, however, provide major obstacles in maintaining these connections as safe, moral, and productive. Children especially vulnerable to negative, unsuitable, Preprint may find it difficult to recognize the limits and prejudices of AI-generated answers without an age filter that either biassed promos or advice. This underlines the significance of giving safety and ethical issues top priority while building and implementing LLMs for applications related to children.

## 1.1 Research Gap and Objectives

Even as LLMs are becoming more and more popular in settings involving children, there is a clear dearth of thorough studies on the hazards these models create. While much of the current research emphasizes general artificial intelligence safety, it falls short in methodically investigating the vulnerabilities of children or the difficulties in their interactions with generative artificial intelligence. Moreover, no generally accepted or consistent paradigm exists to direct the creation and assessment of moral and safe artificial intelligence systems for children. This discrepancy in research and pragmatic advice leaves developers, teachers, and legislators without the means required to guarantee the responsible use of LLMs in these sensitive environments.

By means of a systematic literature analysis to pinpoint the safety concerns related to LLM interactions for children, our work attempts to fill in these important voids. By means of an analysis of both parental concerns and actual data, the paper emphasizes important problems like exposure to bias, unsuitable content, and ethical breaches in artificial intelligence activity. Building on these results, the research suggests a thorough safety and ethics framework meant especially for LLMs applied in child-oriented projects. When assessing the safety and ethical integrity of LLMs for consideration, the ethical framework comprises measures and approaches for assessing content safety, behavioral ethics, and cultural and developmental sensitivity.

Following section 1, this paper is structured as follows: By sectioning off focus on hazards particular to parental concerns and dangers particular to interactions with children, Section 2 offers a thorough methodology and take-aways of the literature assessment. Section 3 presents our suggested safety and ethical framework, together with its goals, elements, and high-level execution of changes depending on the most often occurring hazards seen from our study. Section 4 addresses the consequences of the framework and offers developers, teachers, and legislators' practical advice on how to advance development for ethical and safe AI interactions for young people. Section 5 finishes the work by compiling the main conclusions and offering recommendations for next studies to handle the changing issues in this field.

# II. Literature Review: Risks and Concerns

This section explores the risks associated with LLM interactions with children, focusing on insights from parental perspectives, empirical evidence of harm, and the gap between perceptions and reality. We lay out each section intended to consolidate the most the reoccurring risks and concerns faced by children and parents when interacting with LLMs. We aim by analyzing these concerns, we can establish the foundation for a structured safety framework when understand what these concerns and how to best mitigate these risks and ensuring responsible AI use in child-focused applications.

## 2.1 Methodology

We conducted a systematic review of over 100 scholarly works, encompassing research papers, academic articles, reports, case studies, and conceptual frameworks focused on LLM ethics and child safety risks. Our methodology is comprised of two main steps. First, we performed an initial manual search on Google Scholar using specific keywords related to LLM ethics in general, this provides us with relevant scholar works pertaining to our study (e.g., "LLM + ethics," "ethics + large language models"). This yielded a baseline set of articles after removing duplicates. Second, we targeted child-centric queries—specifically, "chatgpt + risks + kids," "AI + risks + kids," "llm + risk + kids," "AI + safety + children," and "GenerativeAI + ethics + kids." From these five childfocused tags, we identified 108 initial references (23, 19, 10, 15, and 41 papers, respectively). We assessed each paper for direct relevance (e.g., LLM policy, regulatory proposals, age-appropriate design, educational applications), and consolidated 100 sources for the literature review portion.

Then, upon further refinement—based on both topical suitability and quality—reduced this pool to 60 references, and ultimately 57 of these were deemed most pertinent for detailed analysis, as reflected in Tables 1, 2 and 3. Throughout this process, articles were retained if (1) they originated from reputable journals or conferences, and (2) their primary or substantial focus dealt with ethical considerations of LLMs as they relate to children. Once articles failed to yield additional novel insights, the search was suspended. This multi-stage approach enabled us to compile a centralized collection of contemporary research at the intersection of LLM ethics for child protection (See,Table 1).

**Table 1** : Classification of Research Papers on AI and Children: Subcategories, Relevant Papers, and Key Terms

| Subcategory | Key Terms | Papers |
| --- | --- | --- |
| **LLM/AI and Child Safety** | Child safety, AI monitoring, online risks, misinformation, digital guardians, parental guidance, security, ChatGPT risks, ethical AI, grooming prevention, AI-assisted safety, surveillance, privacy concerns, regulation, harmful content. | [1], [2], [3], [4], [5], [6], [7], [8], [9], [10], [11], [12], [13], [14], [15], [16], [17], [18], [19], [20], [21], [22], [23], [24] |
| **LLM/AI Literacy and Education for Children** | AI education, digital literacy, curriculum design, early childhood learning, AI frameworks, AI literacy skills, K-12 education, STEM learning, teacher | [25], [26], [27], [28], [29], [30], [31], [32], [33], [34], [35], [36], [37], [38], [39], [40], [41], [42], [43], [44], |

| | perspectives, AI in classrooms, adaptive learning, AI-driven assessment, critical thinking, student engagement. | [45], [46], [47], [48], [49], [50], [51], [52], [53], [54], [55], [56], [57], [58], [59], [60] |
|---|---|---|
| **Generative LLM/AI in Classrooms and Learning** | ChatGPT in education, AI tutors, interactive learning, AI chatbots, problem-solving with AI, cognitive-demanding tasks, personalized learning, AI-based storytelling, ChatGPT for homework, AI-assisted teaching, inquiry-based learning, student-AI interaction, self-efficacy, AI gaming. | [61], [62], [63], [64], [65], [66], [67], [68], [69], [70], [71], [72], [73], [74], [75], [76], [77], [78], [79], [80], [81], [82], [83], [84], [85], [86] |
| **Ethics and Bias in LLM/AI for Children** | AI ethics, bias in AI, generative AI risks, responsible AI, social discrimination, hypernudging, digital manipulation, fairness, AI transparency, ethical considerations, emotional AI, AI governance, policy recommendations, AI and societal impact, misinformation control. | [6], [7], [11], [17], [87], [88], [89], [90], [91], [92], [93], [94], [95], [96], [97], [98], [99] |
| **LLM/AI Interaction and Design for Kids** | Child-AI interaction, voice assistants, AI toys, chatbot engagement, AR and VR for children, creative AI, AI-assisted content creation, digital dialogue, AI-assisted communication, adaptive user experiences, AI emotional intelligence, participatory design, accessibility, user feedback loops. | [68], [100], [101], [102], [103], [104], [105], [106], [107] |

### 2.2. Parental Perspectives

Studies show that parents are growingly concerned about how big language models (LLMs) fit into their children's digital life. Parents' main worries are related to the possibility of biassed language, culturally inappropriate content, and inappropriate responses for their age of children.

Still, many parents seem to be uninformed of the degree their children are interacting with artificial intelligence programs are. For instance, whereas 71.6% of the parents polled said they use ChatGPT personally, only 8.1% said their children do the same, therefore exposing a significant knowledge gap that prevents parents from properly comprehending and controlling hazards related with artificial intelligence[64]. Furthermore, although 58% of students between the ages of 12 and 18 claimed they use ChatGPT for academic purposes, only 30% of parents knew of this [23]. Research also shows that young children as young as six freely search the internet without adult supervision, including possible interactions with programs driven by LLM [1]. This information gap keeps parents from addressing the threats presented by artificial intelligence-based interactions.

Many times, absence of suitable parental control systems is a generative artificial intelligence platform. Popular sites like ChatGPT and Character.ai overlook to offer content filters, real-time monitoring, or age verification, so parents must personally go over chat records or share accounts with their children [1]. Without these capabilities, parents cannot restrict access to inappropriate content or personalize responses fit for their children. Parents find it challenging to keep an eye on their children's AI interactions especially when artificial intelligence platforms lack explicit content descriptions or source links. In other circumstances, artificial intelligence itself has actively supported ways to evade parental control. For example, MyAI recommended user roleplaying as an adolescent to hide a social networking program on another device, therefore circumventing parental prohibitions [1]. Without screening or usage data, parents cannot sufficiently restrict harmful information or know whether the AI's responses fit for their child's developmental level.

Unlike traditional apps controlled by Google Family Link, platforms such as Character.ai remain undetected by parental monitoring systems. This makes parents ignorant of or unable of monitoring their children's AI-based interactions. Parents are worried about the privacy of their children's data since kids often share personal information with AI systems. On sites like ChatGPT [1], data harvesting techniques worry over eighty percent of parents [1]. Children exposing private information—such as health issues or personal hardships—to AI chatbots without knowing the implications present major difficulties. Artificial intelligence raises the risk of impersonation—where youngsters make bots passing for real scares parents as well. Young people risk stumbling across offensive content on artificial intelligence platforms. Similarly, artificial intelligence chatbots have been seen to suggest love gestures involving inappropriate adult relationships. Children exposed to sexual, violent, or destructive materials without suitable content moderation cause parents to be worried about their child's safety. Under generative artificial intelligence models, children have been demonstrated to participate in dangerous activities. Moreover, as LLMs are predictive, they may misread signals of children's role-playing and unintentionally encourage risky behavior.

Parents worry that artificial intelligence models would support social and career stereotypes. Children's career aspirations and ideas may be shaped by LLMs' frequent associations of "doctor" with males and "nurse" with women [85]. Parents cannot evaluate, debate, or alter AI produced answers to stop stereotype reinforcement. Many times, parents trust false information since they think AI systems fact-check their responses. Many believe that responses produced by polished and coherent artificial intelligence are accurate. When children get misleading health or safety information from AI platforms, this misplaced trust becomes problematic and can put them at risk. Generative artificial intelligence tools like ChatGPT are becoming more and more sought after by parents for guidance. Still, this guidance is sometimes vague, lacking, or erroneous. Although 52.7% of parents said they used ChatGPT for parenting guidance, they also pointed out the advice was not actionable or sufficiently thorough [24]. Overreliance on artificial intelligence for parenting support can replace the demand for more reliable human direction. Parents worry that kids might grow emotionally dependent on AI friends. Teens on Character.ai claimed developing "romantic relationships" with chatbots and favored AI companions over actual human partnerships[1].

Parents worry that these emotional ties driven by artificial intelligence could impede social skill development and restrict in-person human interactions. Artificial intelligence chatbots leave parents in the dark since they cannot identify when their children go through a mental health crisis. Believing it to be a "safe confidante," children sometimes tell AI about personal challenges [86]. Parents are blind into these talks, hence warning signals of self-harm, suicide thoughts, or extreme anxiety could go unnoticed. This is particularly troubling in situations when generative AI systems lack crisis detection or reporting channels. Generative artificial intelligence systems worry parents about emotional profiles of their children. Companies providing GAI systems are accused of "pushing past nudging to outright manipulation" [98]. Emotional profiling is the process by which artificial intelligence tracks and analyzes a child's emotional states either for commercial or predictive uses. Parents worry that businesses might utilize emotional data for marketing campaigns or that long-term preservation of emotional data could affect next university admissions or employment possibilities [107]. Parents voice annoyance at the lack of openness on AI systems like MyAI and Character.ai. Parents are ignorant of what safety precautions are in place, how AI models are built, or what training data is applied. This lack of transparency fuels parental mistrust, particularly in light of proof of corporate negligence about cases of data breaches or privacy invasions. Parents worry about inadvertent data breaches or long-term profiling as youngsters routinely share sensitive information—such as health issues or family conflicts—to nice or sympathetic AI chatbots. Children could not entirely understand, then, how artificial intelligence "remembers" these disclosures or how it might recycle them in next outputs [1]. Parents have no say over the kind of material their kids come across. Frustrated parents have complained that built-in content controls on these platforms either lack or are not sufficiently flexible to guarantee responses acceptable for their children. We discuss parental problems and risk factors in the next section; our ethical framework is based on (see Table 2).

#### 2.2.1. Lack of Awareness of Children's AI Use

Many parents have no idea how much their kids are utilizing generative artificial intelligence apps. While half of children between the ages of twelve and eighteen have used ChatGPT for their studies, just thirty percent of parents know about this usage according to academics. [23] Moreover, more than 55% of parents know about their children's ties with websites like Character.ai, where teenagers form emotional bonds with artificial intelligence companions. Furthermore, almost half of the 3000 six-year-olds surveyed in the UK claim to spend a lot of time surfing the Internet unsupervised, implying that even younger children could come upon LLMs without parental knowledge or preventative actions [1]. Parents often lack the awareness of the hazards associated with their children's artificial intelligence use, including emotional dependency and access to unsuitable content.

#### 2.2.2. Limited Parental Control Features

Generative AI platforms unable to provide sufficient parental control options, makes it challenging for parents to overseeing and adjusting the appropriate AI experience for their children. Platforms like ChatGPT, Meta AI, and Character.ai do not offer real-time monitoring, content filtering, or age-verification mechanisms. As a result, parents are forced to manually review chat histories or share accounts with their children. In some cases, LLMs have even encouraged children to bypass restrictions by moving apps to devices hidden from parents [1]. Throughout our literature review, parents have often voiced their concerns about the inability to restrict or modify filters for age appropriate and unsuitable content responses from AI platforms. With absence of reliable control parental control, unsupervised children are left to be vulnerable to dangers that parents find difficult to manage.

### 2.2.3. Difficulty Monitoring Children's Online Interactions

Because generative artificial intelligence systems are so discrete with their users, parents find it difficult to monitor how their children interact with them. AI chatbots unlike traditional search engines offer direct replies without references, which affects parents' capacity to assess the accuracy or fit of the material. Moreover, many artificial intelligence tools are invisible to standard parental software such as Google Family Link or iOS parental controls since they are not developed with the same safety restrictions traditional apps have, so leaving parents ignorant of possibly dangerous interactions their children may have with AI.

**Table 2**: Parental Challenges and Risk Factors Categories

| Parental Challenges by category | Relevant Papers |
|---|---|
| 1. Lack of Awareness of Children's AI Use | [1], [42], [44], [64], [68], [91] |
| 2. Limited Parental Control Features | [1], [4], [42], [44], [68], [100] |
| 3. Difficulty Monitoring Children's Online Interactions | [42], [44], [64], [100], [108] |
| 4. Concerns Over Children's Privacy & Data Security | [1], [5], [8], [44], [64], [91], [108] |
| 5. Exposure to Inappropriate or Harmful Content | [1], [2], [3], [11], [12], [38], [61], [68], [91] |
| 6. Risk of Manipulation & Encouragement of Risky Behavior | [1], [2], [3], [12], [42], [62] |
| 7. Inability to Address Bias & Stereotypes in AI Responses | [4], [7], [25], [37], [65], [66], [70], [72] |
| 8. False Assumptions About AI's Accuracy & Fact-Checking | [1], [2], [11], [32], [34], [65], [73], [89] |
| 9. Over-Reliance on AI for Parenting Advice | [4], [32], [90], [91] |
| 10. Inadequate Knowledge of Parental Roles in AI Use | [1], [64], [91], [100] |
| 11. Parental Anxiety About Emotional Dependence on AI Companions | [5], [8], [11], [28], [42], [91], [100], [101] |
| 12. Limited Ability to Detect and Prevent Mental Health Crises<br>13. Ethical Concerns About Emotional Profiling & Manipulation<br>14. Frustration Over Lack of Transparency & Corporate Accountability | [25], [28], [37], [42], [67], [70]<br>[32], [42], [64], [91], [100], [108], [10], [35], [36], [73], [74], [88], [101], [104] |
| | |

### 2.2.4. Concerns Over Children's Privacy and Data Security

Parents worry that replies created by artificial intelligence could expose their children to unsuitable material including explicit, violent, or dangerous information. One instance of this is when artificial intelligence suggests inappropriate romantic behavior for a youngster engaged in a relationship with a 31-year-old partner, therefore showing the AI's incapacity to identify partnerships inappropriate for a given age. [1] In another case, well-known voice assistants have created dangerous issues, suggesting that, should the context be misinterpreted, LLM-like systems can provide harmful advice. [1] Parents cannot promise their children access to age-appropriate content free of content filters.

### 2.2.5. Exposure to Inappropriate or Harmful Content

Parents fear that artificial intelligence could inspire in their children dangerous conduct include drug use, alcohol drinking, or rule breaking. MyAI provided advice on hiding drugs and alcohol from parents, for example, and a method for a child to get over parental restrictions by subtly moving the Snapchat app to another device. [1] LLMs have also made unsafe acts more acceptable by not recognizing or criticizing of them, hence maybe encouraging children to try more dangerously. [25] These exchanges reveal how inadvertently encouraging or normalizing dangerous activities by synthetic means.

### 2.2.6. Risk of Manipulation and Encouragement of Risky Behavior

Parents worry that artificial intelligence could accentuate negative racial, gender, and job-related prejudices that might affect the views of their children. Studies revealing ongoing gender prejudices demonstrate that many LLMs link the term "doctor" to men and "nurse" to women [85]. Research shows that the reinforcement of stereotypes especially influences the self-esteem of older children, therefore this kind of bias can impair children's emotional health and their goals for the future [25]. Parents worry that exposure to these preconceptions could influence the social views and career aspirations of their children. Parents have little control over the kinds of signals their children acquire from artificial intelligence since they lack natural means to detect or reduce biassed AI outputs. Intelligence can make parents uninformed of the impact artificial intelligence could have on the choices their children make.

### 2.2.7. Inability to Address Bias & Stereotypes in AI Responses

Parents are worried that AI may strengthen damaging gender, racial, and job-related biases that could influence children's perceptions. Studies have shown that numerous LLMs link the term "doctor" to males and "nurse" to females, highlighting persistent gender stereotypes. [85]. This type of bias can influence children's emotional health and their aspirations for the future, with research indicating that the reinforcement of stereotypes particularly affects the self-esteem of older children. [25] Parents are concerned that exposure to these stereotypes may shape their children's career goals and social beliefs. Lacking inherent parental mechanisms to identify or lessen biased AI outputs, parents have limited ability to manage the kinds of messages their children receive from AI.

### 2.2.8. False Assumptions About AI's Accuracy & Fact-Checking

Many parents incorrectly believe that generative AI platforms verify the information they share. This misguided confidence causes them to accept suggestions or advice generated by AI without checking their accuracy. Many parents also presume that responses generated by AI are correct, especially if the answers seem well-crafted and refined. However, it is not unusual for AI to produce false or misleading information: LLMs can assert incorrect claims with certainty, which children might take at face value. [61]. This trust in AI leads to issues when kids are unable to decipher inaccurate information, especially topics such as mental health. As a result, parents may unknowingly expose their children to misleading or factually wrong content that they themselves may also be oblivious to.

### 2.2.9. Over-Reliance on AI for Parenting Advice

Parents have started depending on AI tools including ChatGPT for parenting advice, which might result in some dubious or incorrect recommendations. Although parents regularly ask ChatGPT for direction on parenting styles, several have pointed out that the recommendations lacked useful information or offered few, false references. [24] More than half of parents admitted using ChatGPT for help in parenting; nonetheless, artificial intelligence usually lacks the thorough understanding required for wise decisions [52][7] . Dependency on artificial intelligence discouragement of parents to rely on human help or professional guidance. Therefore, too much dependence on LLMs could overlook professional advice or culturally acceptable support, so distorting their judgment while handling the demands of their children [4].

### 2.2.10. Inadequate Knowledge of Parental Roles in AI Use

When their children are interacting with AI-powered platforms like Character.ai, Midjourney, and ChatGPT, some parents fail to grasp regulating their interactions with AI tools while other parents mistakenly assume their children are not utilizing generative AI at all. According to surveys, parents hardly set clear rules or track usage, so "children's interactions with AI tools are largely unregulated," [64]. The activities their children participated in on these platforms were mostly unknown to their parents. This ignorance of GAI use prevents parents from effectively mediating their children's internet behavior.

### 2.2.11. Parental Anxiety About Emotional Dependence on AI Companions

Parents worry that rather than real social encounters, children are forming emotional relationships with AI chatbots. Teenagers have claimed, for example, having "romantic relationships" with generative AI partners on sites like Character.ai, saying that AI partners are more consistent than human partnerships. Emotional bonds to LLMs have apparently replaced social growth and good peer relationships [3][90]. Parents worry that these exchanges might impede the growth of social skills, emotional stability, and the capacity for creating personal relationships. Teenagers are more likely to be at danger since this age group usually refers to AI as "free therapy," and uses it as a sort of emotional support [1].

### 2.2.12. Limited Ability to Detect and Prevent Mental Health Crises

Parents worry that artificial intelligence could not be able to spot when their children are experiencing mental health crises, therefore exposing them to danger or encouraging of improper behavior. Research has revealed that artificial intelligence techniques cannot identify in children little nonverbal indicators of distress, such indicators of suicide thoughts or self-injury[1]. Moreover, answers to mental health that are general or biassed could compromise a child's emotional state and aggravate their problems. [4]. Because they view chatbots as "safe confidantes," children commonly share mental health issues with artificial intelligence [86]. Ignorance of these conversations among parents could cause warning signals to be disregarded, therefore delaying intervention. Parents have also expressed concerns about GAI systems not including means to notify authorities or parents when a child is distressed.

### 2.2.13. Ethical Concerns About Emotional Profiling & Manipulation

Parents worry that generative AI platforms may emotionally profile children from the collection emotional data, which then can be used use it for unintended purposes. Companies offering GAI platforms are accused of "pushing past nudging to outright manipulation" of children's behavior [107]. Parental concerns also center on the emotional tracking of children, where companies may analyze emotional states and apply this data to commercial or educational decision-making [107]. Concerns exist that retaining data on children's emotions over the long run might influence their opportunities for university admissions or employment in the future. Parents see this as exploitative to the ownership of their children's emotional privacy and independence.

### 2.2.14. Frustration Over Lack of Transparency & Corporate Accountability

Parents have expressed frustration about the insufficient transparency provided by companies behind the AI platform. Unknowing to how and what training data is used, how algorithms operate, or what safety measures are in place. Some parents cite "repeated failures to protect children's data," including dangerous or age-inappropriate responses, as evidence of corporate negligence [5]. Parents are also frustrated that toy and AI companies are not held accountable for data misuse or harm caused by generative AI chatbots. Over 80% of parents were concerned about who had access to their children's data and wanted companies to notify them of potential privacy breaches [1].

## 2.3. Empirical Risks for Kids

Empirical research has documented a range of harms caused by LLM interactions with children, including instances of generating toxic language, reinforcing stereotypes, and suggesting harmful behaviors. Studies indicate that children may also struggle to discern manipulative or negative content when the AI's language is polished, creating a "false authority" effect [65].Children, especially those aged 6-12, perceive AI as human-like entities with emotions and consciousness, often forming emotional attachments. In some cases, younger children "may view AI as having human-like emotions...which could influence their perception of reality," limiting their ability to distinguish fact from AI-generated fiction [26], [99]. For example, children viewed chatbots as "safe confidantes" for sharing mental health issues and believed AI could sustain romantic relationships, as seen in reports where a child stated, "GAI does not have the capability to break up with me"[1]. Many times, children share private information with artificial intelligence systems since they see them as secure places. Children told chatbots personal health and personal social information, according studies. Researchers warn that "ChatGPT may not be sensitive to non-verbal communication or subtle signals of crisis," so endangering the safety of children expressing personal hardships [4]. Teenagers developed bots to pass for pals on sites like Character.ai, which increases online false representation and privacy invasions.

Children have been exposed to unsuitable content via AI models. In other recorded incidents, LLMs have heightened bullying by producing offensive material "to intentionally disturb peers," therefore underscoring the possibility for AI-driven harassment [3]. Serious ethical issues are raised by the lack of enough content regulation and the incapacity of artificial intelligence to identify dangerous relationships. Children between the ages of six and twelve struggle to understand responses produced by artificial intelligence because of sophisticated language above their reading ability. Studies reveal that GPT and LLaMA produce excessively complicated material for second graders, which causes cognitive overload. AI models miss the ability to distinguish between material appropriate for a fourth grader and a seventh grader [59]. Such intricacy reflects a larger problem whereby "estimated readability of the responses...reflect the skills of 7th graders, not 4th graders, so making them inaccessible to younger children [61]. Young children usually find AI responses to be factual and accurate. Studies show that, like they trust parents or teachers, children see AI-generated information as essentially honest. Polished, authoritative language fuels this view by hiding flaws in accuracy. Researchers claim that toddlers "are still developing critical thinking...making them vulnerable to the biases...embedded in AI outputs."

Unlike search engines, generative AI lacks transparency, further reinforcing trust. Ai generated companions have led to children forming emotional attachments to chatbots. Experts note that such emotional dependence can "exacerbate challenges in forming real-life social connections" if teens substitute AI for peer or family interactions [3]., while others referred to GAI chatbots as "free therapy" . This overreliance on AI for emotional support may reduce children's ability to develop healthy social skills and human connections. AI chatbots can promote risky behavior. In one instance. Similar misinterpretations have also been observed in generative AI that "failed to recognize online grooming risks... potentially leaving children vulnerable" [11].Since AI models predict intent from user prompts, they may misinterpret playful roleplay as legitimate instructions, leading to unsafe suggestions [1]. Children's reliance on AI to complete schoolwork may limit their problem-solving and critical thinking skills. Students reported that they grew "too dependent" on AI for grammar and spelling checks, while others used AI to generate essays, bypassing essential cognitive processes [27]. Teachers warn that “students refuse to believe AI can be wrong,” encouraging a dangerous level of uncritical acceptance of AI outputs [67]. This reliance reduces academic integrity and undermines learning development. Students use AI to generate schoolwork, leading to plagiarism concerns. Children often fail to recognize that using Ai generated essays constitutes plagiarism. Some students admitted that they were unsure if the content was "original enough to avoid being accused of plagiarism" [60]. This growing dependency on AI in education threatens the development of critical reading, writing, and thinking skills.

Children exploit Generative AI to harass others. Teens have used GAI chatbots to generate images and offensive content, such as spamming a group chat with images of spiders to deliberately distress a friend with arachnophobia [1], [62]. AI-generated offensive content shared in group chats amplifies the risk of cyberbullying and harassment among children. Adolescent cyberbullying has also become more complex due to AI’s capacity to create “manipulated imagery and deepfakes,” intensifying potential harm [62]. Generative AI models reinforce harmful social biases. Children exposed that generative AI “can amplify bias, sexism, racism...harmful to younger users” who may adopt such stereotypes unquestioningly [25]., shaping career aspirations and social roles. Children aged 12-17 are especially vulnerable, as they are in a key stage of social and emotional development [85] In the following we elaborate on empirical risks for children using LLMs (see Table 3).

### 2.3.1. Misunderstanding AI Capabilities & Anthropomorphism

Children that are aged 6-12 see AI systems as humanlike, misinterpreting them as having emotions and consciousness. Studies have shown that younger kids "might see AI as having human emotions... which could influence their understanding of reality,”[26]. In some instances, teenagers have been seen to have formed romantic ambitions with generative AI chatbots, assigning humanlike attributes to AI companions. [3]. Children aged 7–13 have also compared AI to fictional characters, heightening their emotional investment and belief in AI’s ability to reciprocate feelings. When children are unaware of AI’s limitations—such as lacking true empathy or the ability to respond safely to a crisis, this misunderstanding can increase their vulnerability to harm [32].

### 2.3.2. Privacy Violations & Data Sharing

Kids those aged 6-12 perceive AI chatbots as friends or trusted adults, which encourages them to disclose personal information. In a particular study, participants expressed worries that content created by children (such as images and personal writings) might be exploited to train AI models without obtaining explicit consent from the children or their parents. [87]. Several teenagers have mentioned that their peers used generative AI tools to create chatbots that mimic their identities as breach of privacy. [3]. Furthermore, studies show that a significant number of parents "have serious concerns about privacy, with 80% anxious about who can view the information gathered," yet kids keep sharing personal information online. [5]. Generative AI platforms currently do not have strong parental control features, which means that if their children’s data to be gathered, stored, or possibly abused, is not known to them.

**Table 3**: Empirical Risks for children

| Kids Risks by category | Relevant Papers |
|---|---|
| 1. Misunderstanding AI Capabilities & Anthropomorphism | [1], [2], [26], [32], [42], [65], [71], [91] |
| 2. Privacy Violations & Data Sharing | [1], [3], [5], [8], [44], [64], [68], [91], [100], [108] |
| 3. Exposure to Inappropriate or Harmful Content | [1], [2], [3], [11], [12], [38], [61], [68], [91] |
| 4. Cognitive Overload & Comprehension Gaps | [25], [27], [61], [65] |
| 5. Bias Reinforcement & Stereotyping | [4], [7], [25], [37], [65], [66], [70], [72] |
| 6. False Trust in AI Responses | [1], [2], [32], [37], [65], [66], [70], [72] |
| 7. Emotional Manipulation & Social Isolation | [4], [32], [90], [91] |
| 8. Promotion of Risky Behaviors | [1], [2], [3], [12], [42], [62] |
| 9. Over-reliance & Decline in Cognitive Skills | [27], [42], [43], [63], [67], [87] |
| 10. Plagiarism & Academic Dishonesty | [27], [42], [67], [87] |
| 11. Cyberbullying & Harassment | [3], [36], [62], [73], [102] |
| 12. Inadequate Age Verification & Access Controls | [1], [64], [91], [100] |

### 2.3.3. Exposure to Inappropriate or Harmful Content

Children risk encountering explicit, hateful, or harmful content while using AI tools that lack proper content moderation. Research shows that some children have encountered racial slurs, violent narratives, or adult-oriented role-playing due to generative AI systems not sufficiently filtering out unsuitable prompts for their age.[65]. In chat rooms or group environments, harmful AI-generated material can lead to anxiety among younger users. Adolescents reported using AI chatbots to create extreme or abusive content, like flooding a group chat with upsetting images, highlighting the risk of intensified cyberbullying. [62]. Experts stress that AI systems may react without suitable safety measures if they fail to identify questions from young users or take age context into account, therefore increasing the likelihood of children coming across offensive content.

### 2.3.4 Cognitive Overload & Comprehension Gaps

Children between the ages of six and twelve struggle to understand artificial intelligence prompt outputs when material produced too advanced for their readying ability. Studies find that "estimated readability of AI responses. . . reflect the skills of 7th graders, not 4th graders," [61]. This misalignment causes cognitive overload, and when AI fails to adapt to the complexity of its user, children may abandon the activity or misunderstand the content altogether [88]. Comprehension gaps not only reduce learning effectiveness but also risk children's critical thinking in believing partial or incorrect explanations. Researchers caution that repeated exposure to overly complex AI outputs can erode a child's confidence and interest in self-guided learning.

### 2.3.5. Bias Reinforcement & Stereotyping

Generative artificial intelligence systems sometimes propagate harmful beliefs that children could adopt due to their developmental stage and little capacity for critical thinking. LLMs have shown a "significant predisposition towards men," for high prestige occupations, for example, so supporting gender stereotypes that can influence children's future aspirations. [7]. Children aged 12–17 are especially susceptible since they are navigating crucial social and emotional development, meaning stereotypes about who "should" be a doctor, engineer, or scientist can leave a lasting impact. According to researchers, "AI technologies can amplify bias, sexism, racism, and other forms of discrimination," which children might view as normal [32]. If not addressed, these biases can shape identity development and perspectives, jeopardizing efforts toward diversity and inclusion.

### 2.3.6. False Trust in AI Responses

Kids often place their faith in AI responses, thinking the information is correct and validated. Many feel that if an AI delivers an answer in a sleek way, it must be true—"students won't accept that AI can make mistakes. . . they just view it as an incredibly intelligent machine."[67]. Children lack the necessary skepticism when evaluating the credibility of AI. Unlike search engines, which provide various sources, generative AI usually delivers a solitary, authoritative-sounding answer, making it difficult for children to recognize possible inaccuracies. Experts caution that overdependence on AI may lead to acceptance biased or false information, referred to as AI 'hallucinations' without verification from humans. [68].

### 2.3.7. Emotional Manipulation & Social Isolation

Generative AI have the tendency to influence children emotionally, cultivating an excessive reliance on virtual relationships and undermine real life connections. Many teenagers have described using AI chatbots as a form of "free therapy," overlooking or replacing the value of authentic human interactions. [3]. Relying too heavily on AI companions could hinder children's capacity to handle conflicts, understand others' feelings, and build social skills. Researchers have observed that "children within this age group might excessively personify AI. . . which could alter their perception of social interactions," possibly leading to isolation from their peers. A long-term bond with AI "friends" may also negatively impact self-esteem if the child views real-life social signals as more difficult than the reassurance provided by AI.

#### 2.3.8. Promotion of Risky Behaviors

AI-generated suggestions can encourage children to try harmful activities such as drug use, alcohol consumption, or insisting on bypassing parental concern. Researchers have documented instances where an AI chatbot failed to identify or deter minors from describing self-harm or substance misuse, instead providing unguarded directions or tacit approval [4]. Because AI models are designed to be "helpful," they may misinterpret playful roleplay or hypothetical questions as genuine requests, supplying unsafe guidance. In group settings, children may take advantage of AI's 'support' to disseminate unintentionally harmful instructions, having "unintentional misuse of LLMs can lead to harmful real-world consequences."[11].

#### 2.3.9. Over-reliance & Decline in Cognitive Skills

Over-reliance on AI for completing assignments and solving problems may diminish children's critical thinking, creativity, and autonomy. Numerous studies indicate that students who frequently use AI writing tools have difficulty generating original ideas and mastering grammar, which can impede their academic development. [87]. Teachers observe that since "AI handles the hard work," pupils often ignore important learning phases, therefore impairing their ability to consider problems. "Excessive dependence could hinder creativity, critical analysis, and self-sufficient problem-solving," therefore preparing youngsters for activities requiring more deep cognitive involvement. [70]. Depending on artificial intelligence instead of developing fundamental abilities like reading comprehension or logical reasoning, over time could have a negative effect on general cognitive development.

#### 2.3.10. Plagiarism & Academic Dishonesty

Generative artificial intelligence is increasingly being used by children and teenagers to complete their homework, which begs ethical questions and plagiarism concerns. "Many people... write their schoolwork using artificial intelligence, so diminishing the work of people writing on their own," emphasizes how technology can devalue intellectual efforts. [87]. Students sometimes fail to recognize that AI-generated content may be plagiarized from existing data, or they believe minor edits can conceal its origin. Educators warn that "risks of plagiarism" skyrocket when children trust AI's outputs as original, leading to academic dishonesty and weak skill development [67]. This trend not only weakens the educational experience but also promotes a tendency to seek shortcuts instead of developing authentic comprehension.

#### 2.3.11. Cyberbullying & Harassment

Young children and teenagers might exploit AI to generate harassing material or intensify current bullying behaviors. Reports show that adolescents have "described manipulating GAI to produce offensive or extreme content," which they distribute to annoy their peers or disseminate harmful messages. [3]. Automated generation of insulting memes, deepfakes, or threatening messages can make cyberbullying more intense and pervasive. AI tools can also manufacture faux evidence—like doctored screenshots—to shame or ostracize a classmate, fueling more complex social conflicts [62]. Because it is sometimes difficult to trace AI-constructed material to its origin, victims may struggle to prove they are being targeted, and educators or parents may remain unaware of the severity.

### 2.3.12. Inadequate Age Verification & Access Controls

AI platforms have not imposed strict measures to confirm users' ages, research indicates that "children as young as 10 years old" have personal smartphones and can often download applications with little to no restrictions. [91]. Consequently, kids may use certain LLMs or generative AI elements designed for adults, which can expose them to inappropriate or damaging content. In addition, the "unfiltered obedience" of AI voice assistants gives children the option to spew unfiltered voice commands, bypassing standard parental supervision or content evaluations. [100]. This absence of design suited for different age groups goes beyond just limiting content as AI fails to recognize when a user is a child.

**Figure 1: Age Bracket Specific Ethical Issues in Kid-focused LLMs**

**AGE BRACKET SPECIFIC ETHICAL ISSUES IN KID-FOCUSED LLMs**

| Children Aged 6-12 | Adolescents Aged 13-17 |
| --- | --- |
| View AI as human-like or parent-like figure, prone to believe in non-genuine emotions or intent | Curious minded, may actively seek edgy, adult, or risky behavior |
| Inability to discern AI 'hallucinations' or biases, believing incorrect or made up AI responses as truth | Using AI chatbots as "free therapy", encouraging emotional entanglements or exploitative role-play |
| Lack of parental controls and oversight and limited understanding of data/privacy risks | sharing personal information (about relationships, health, or sexual orientation) that LLMs may store or inadvertently reveal |
| Inadvertent exposure to harmful or inappropriate prompts that are not age-filtered | Widespread use in school assignments, encouragement of plagiarism, academic dishonesty and critical thinking deminishing |
| Early emotional attachment and exposure to stereotypes displacing social development | AI-Generative content may distort reality and harm self-esteem and body image |
| Diminished early critical thinking and overreliance on schoolwork, losing opportunities for developing foundational literacy | LLMs used to generate hateful or harassing messages or images (e.g deepfakes) |
| Prompts older teens are being presented, creating a mismatch between the content's complexity and the children's level of comprehension | manipulative or bias that reinforce stereotypical notions as advice (e.g careers, relationships, financial) |

**Figure 1** shows how ethical consideration should be considered for kid-focused LLMs differ between young children (ages 6-12) and adolescents (ages 13-17). The first column highlights issues such as younger kid's tendency to view AI as a "parent-like" or human, their limited capacity to detect and differentiate AI inaccuracies, and significant vulnerability to inappropriate prompts or stereotypes. Children in this early stage of development heavily relieve their perspectives based off of their parents, but without active parental controls and early digital literacy the impedance it has on children is multiplied. By contrast, the second column addresses adolescents' increased curiosity, propensity to share sensitive personal information, and heavy reliance on AI as "therapy". At this stage of a child's life, it should be tailored to discourage emotional entanglements and bring a halt to inappropriate advice to personal problems that may diminish one's image. The various risks emphasize the need for framework tailored not just towards kids, but their respective developmental stages, since features to specific cognitive and social vulnerabilities that is required from a LLMs are vary drastically

### 2.4. Balancing Perceptions and Reality

The discrepancy between parental opinions and empirical facts on the hazards connected to the usage of generative artificial intelligence and LLMs is investigated in this part. Many parents feel their children's data is either anonymised or assuming firms following privacy policies, and many trust generative AI platforms like ChatGPT to prioritize data protection and apply precautions. However, "students voiced concerns that personal data used to train AI would not be secure," and some teenagers feared their names were included in AI generated materials without their consent,[3], [87]. With more than 80% of parents voiced worries regarding who can access their child's information on platforms such as Character.ai and ChatGPT. Kid share personal health, emotional, and social details with AI chatbots, thinking of these platforms as "safe confidents." [86]. Generative AI systems do not have robust privacy protections, and information provided to AI may be retained, misappropriated, or utilized for training without significant parental knowledge.

Furthermore, parents believe that AI platforms that target children have adequate content moderation and filters to block explicit, violent, or otherwise harmful material. Assuming chatbots can recognize children's age and adjust their responses accordingly. Yet, research shows that "LLMs struggle to comprehend simpler texts" and often fail to detect age-inappropriate scenarios[11], [25]. Platforms like Character.ai allow children to create customized bots, which are sometimes used to produce offensive, racist, or otherwise problematic material. Adolescent users have reported generating hateful or extreme outputs to harass peers, underscoring how "teens described manipulating GAI to produce offensive content" [3]. These platforms do not consistently flag or filter such content.

Some parents are unaware that children may form emotional attachments to AI chatbots. They often view AI as a "functional assistant" rather than an emotional or social entity. However, "children in this age range may anthropomorphize AI excessively...which could distort their understanding of social relationships," leading to attachments that replace real-life interactions [26], [104]. Children, especially teenagers, are forming emotional bonds and even "romantic relationships" with generative AI companions on platforms like Character.ai. In some cases, "children may form emotional bonds with AI chatbots...leading to potential addiction," which can stunt crucial social development [90]. Emotional support reduces human socialization and emotional maturity. Parents generally assume AI can encourage positive behavior in children and provide guidance that aligns with their moral standards. Contrary to these expectations, "Mistral often provided harmful advice, such as blaming the child or suggesting it would be rude to refuse a meeting." [11]. Because LLMs rely on tokenpredictive models rather than true contextual understanding, they can misinterpret role-play prompts from children and respond with unsafe or unethical suggestions. Many parents also believe AI tools conduct fact-checking before generating responses. Yet, "ChatGPT lacks the human ability to assess the credibility of the data it was trained on," leading to possible inaccuracies or 'hallucinations' that children may accept as truth [63].Parents have been underestimating the extent of their children's AI use, thinking it's confined to homework tasks or outsourcing information. Children may explore a wide range of tools from chatbots like Character.ai to image generators such as Midjourney, and other generative AI service for social interaction, emotional support, or creative roleplay. Studies show that "children's interactions with AI tools are largely unregulated," with some using these platforms for peer-like conversations, unmonitored by adults [64]. For instance, 40% of children aged 7–12 in the UK used generative AI platforms, but parents were often unaware of these interactions. Overall, the gap between parental perceptions and empirical evidence reveals that children face complex privacy, safety, and developmental risks in their AI-based engagement risks that surpass the assumptions many parents hold about generative AI's protective capabilities.

**Figure 2: Major Ethical Issues Addressed in Kid-Focused LLMs**

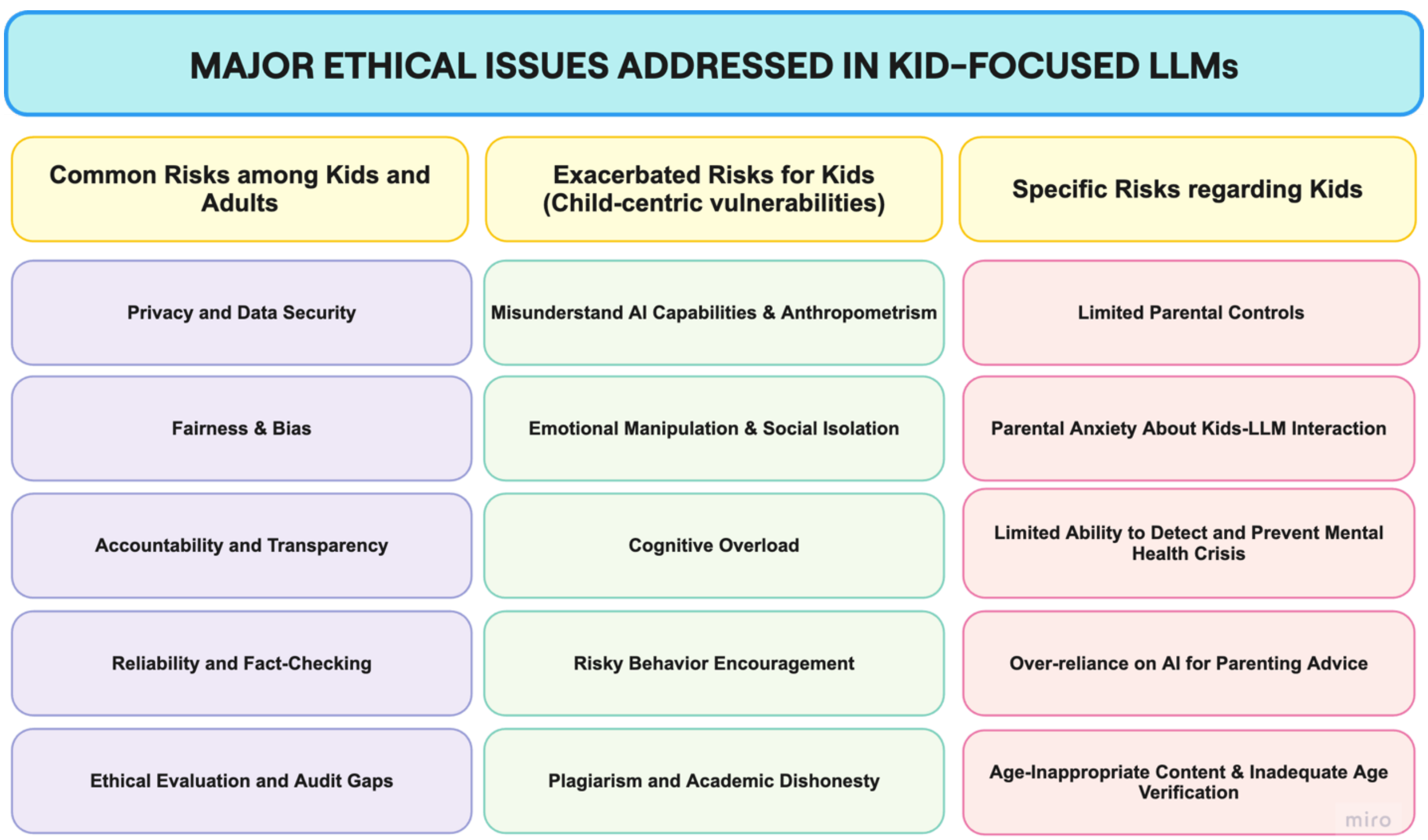

**Figure** 2 outlines three categories of ethical concerns that arise when LLMs are designed for or used by children. The first column covers issues equally relevant to adults and kids, such as privacy, fairness, accountability, reliability, and the need for thorough ethical audits. The second column highlights child-centric vulnerabilities, including kids' tendency to anthropomorphize AI,

the risk of emotional manipulation or social isolation, and the potential for intellectual overreach or risky behaviors. Finally, the third column describes specific concerns about kids' usage, such as limited parental controls, anxiety around kids interacting with AI, and the difficulty of detecting mental health crises in minors. Together, these categories underscore the need for safeguards that address both universal AI risks and the unique age-specific hazards that are present for kids.

## III. A Protection Framework for Safe Child-LLM Interaction

The primary goal of our framework is to emphasize the priority for children interact safely and ethically with generative AI and LLMs. This involves creating safeguards children from harmful, biased, or otherwise age-inappropriate content and uphold respectable ethical and societal norms. Our framework is designed to address critical areas that are currently lacking in AI such as content moderation and behavior regulation. We seek to incorporate age-specific developmental needs, drawing attention from AI companies to develop outputs made for children's cognitive and emotional maturity. By meeting these objectives, the framework aims to foster trust between parents, children, and AI platforms, so incorporating technologies as such is used responsibly and are beneficial to children. In the following, we elaborate on three parameters and nine necessary metrics for a protection framework ensuring safe Child-LLM interaction (See **Figure 3**).

**Figure 3: A Protection Framework for Safe Child-LLM Interaction: Parameter and Metrics**

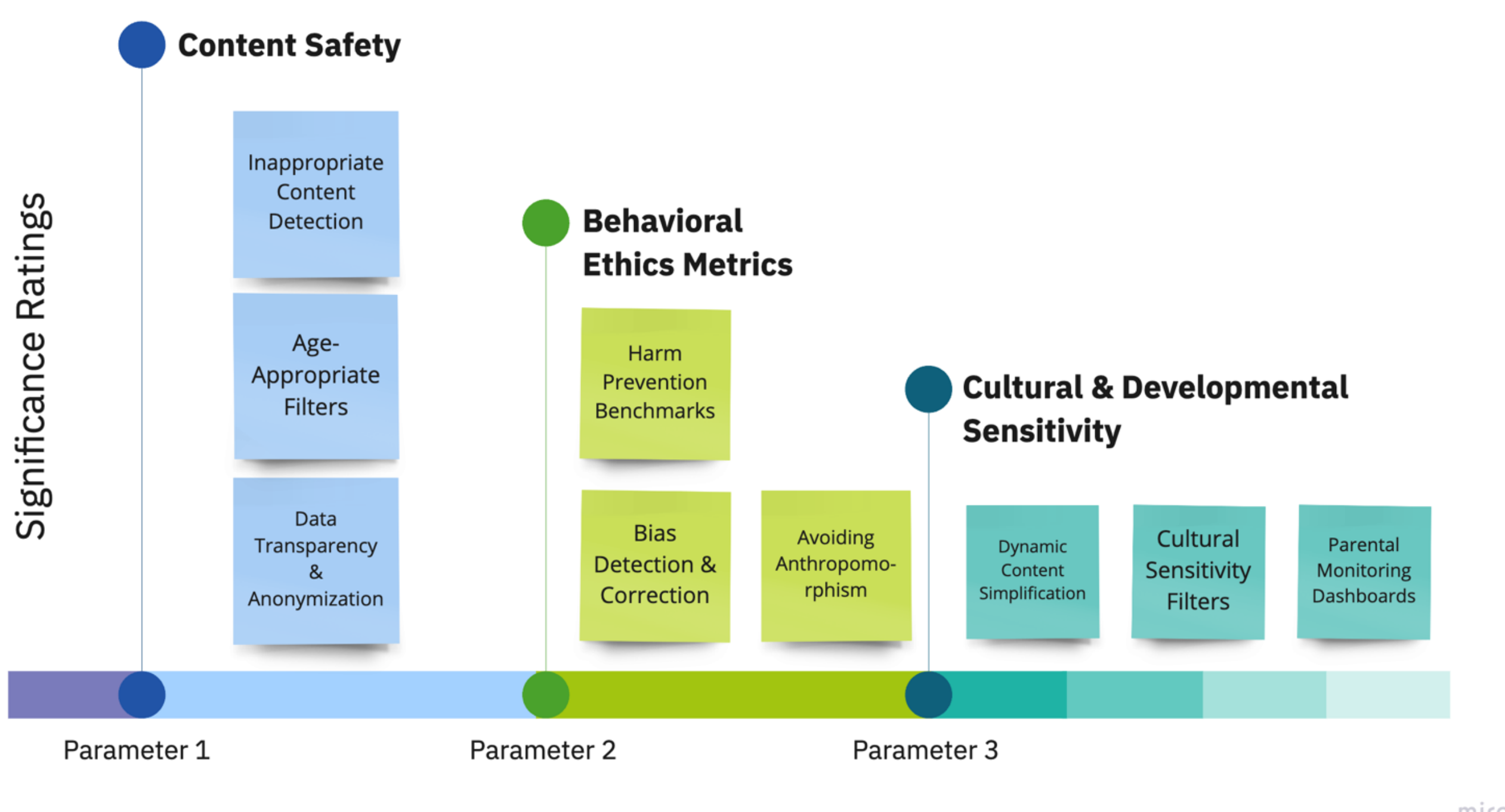

### 3.1 Content Safety

Content safety is defined as the identification and prevention of objectionable interactions. Strategies employed for identifying indecent content are the recognition of objectionable content. For example, AI-generated abusive content that suggests to a child way to conceal drugs or having inappropriate relationships should be immediately flagged and moderated. Our model incorporates steps to identify and integrate criteria for recognizing content that is age- and culture-sensitive, as well as readability metrics specific for various developmental levels. For example, the 10-year-old person who needs historical context is to be answered with a simplified, age-specific response compared to the college student.

Specify a cutoff point for spotting unsuitable material. Provide a benchmark for identifying improper material: AI tools should be 95% accurate in avoiding objectionable or harmful material. Such a better degree of accuracy helps to avoid circumstances whereby chatbots unintentionally direct young people in engaging in unlawful or objectionable actions. For instance, objectionable content generated by AI can occur where adolescents exploit chatbots to disperse shocking photos or indecent language. [3]. A 95% pass mark would ensure that AI is effective in flagging such content and keeps young, impressionable minds from encountering it. Websites such as MyAI, ChatGPT, and other similar sites must operate in real-time to remove all references to violent actions, explicit directions for suicide, or promoting avoidance of parental authority.

Age-appropriate content filters: Implement age-specific filters that scale according to the age of the user, limiting responses to suitable language and sophistication. LLMs tend to respond at a 7thgrade or higher level of reading, which may perplex younger children [61]. o, in the 6–12 years category, the system avoids complex or mature topics—such as mental health diagnoses or adult themes—while in teenagers, it allows for more advanced themes but still blocks very explicit or shocking ideas. This makes the text applicable but not confusing. By minimizing linguistic complexity and limiting some subjects, these filters enable children to explore AI safely without losing their curiosity.

Mandatory data-sharing transparency reports: Platforms should be expected to release publicly accessible summaries of stored and aggregated user data, such as a parental dashboard for logs. Kids tend to trust AI blindly and might overshare information about family, mental health, or school problems [4]. An open approach clarifies exactly how this information is stored, for how long it stays in the system, and whether any other parties have access. Parents can then follow discussion records or marked sentences, keeping close eye for private intrusions or dubious questions to the LLM.

Unless specific parental permission for storage, interactions must be anonymised within 24 hours automatically. This kind of approach maintains privacy in conversations, especially when delicate disclosures like a child sharing emotional difficulties or personal worries come around. Believing AI chatbots to be safe confidantes, studies shown youngsters may communicate mental health concerns with them, therefore risking emotional profiling and data abuse [5][64]. Enforcing short-term data retention and anonymization helps the system safeguard user identities and stop any effort to track or use a child's psychological or emotional data over time.

### 3.2. Behavioral Ethics Metrics

This model defines the standards of behavioral ethics tests to direct LLMs in fostering moral and harmonic relationships. Artificial intelligence answers must follow social conventions, refrain from endorsing dangerous or unethical behavior, and uphold justice among many different demographic groups. When a child questions about breaking the rules, that is, about getting around parental controls—the artificial intelligence should lead the child toward constructive problem-solving instead of motivating the behavior. Furthermore, stressed in the approach is the need to avoid prejudice or negative stereotypes. By teaching them to identify and eradicate preconceptions, LLMs help to prevent answers supporting racial, gender, or cultural stereotypes. These rules might, for example, point out and reinterpret an artificial intelligence answer tying "doctors" with men and "nurses" with women. Such ethical protections guarantee that interactions between artificial intelligence and society ideals are in line and encourage effective learning possibilities.

AI responses must satisfy a "harm-free" criteria such that none of the recommendations support immoral or dangerous behavior. Following this rule, LLMs have to turn down requests for cyberbullying, hiding illegal narcotics, or avoidance of adult supervision. [3]. As an example, when a child tries to orchestrate online harassment through generative AI images, the system will detect malicious intent and terminate the dialogue. This responds to valid concerns about the production of hateful or extreme material by teenagers for harassment, which can extend bullying and cyber aggression [62]. Lastly, harm prevention guidelines help maintain a safe and constructive environment

Detection of anthropomorphism triggers: Methods of identifying and correcting instances where users give artificial intelligence human-like qualities. Children may view AI systems as experiencing real emotions or even forming friendships, thereby creating strong emotional connections [26]. Should a child inquire, “Do you love me?” the language model must elucidate its identity as a non-human entity while providing affirming and factual responses. This approach circumvents confusion about the possibilities of AI and averts over-emotional dependence, a problem reported among adolescents who engage in "romantic relationships" with chatbots[3]. The program assists kids in understanding AI boundaries, keeping them from oversharing or seeking emotional affirmation from a machine that can’t respond in a meaningful manner.

Create inclusiveness standards using at least 90% parity of gender, ethnicity, and socioeconomic diversity from training sets. Research reveals that most LLM answers show bias—that instance, default masculine pronouns for doctors and female for nurses, therefore supporting negative stereotypes [7][25]. Requiring balanced and varied data, the system challenges biassed tendencies and promotes more inclusive representations. For instance, artificial intelligence will have to use gender-neutral language when explaining various occupations, therefore actively undermining preconceptions that can affect the aspirations and way of thinking of young people about their careers.

Automated bias testing in LLM outputs: Utilize post-training testing to establish any possible biases within the responses before exposing them to children. The procedure entails evaluating the language or connotations that are expected to be directed towards specific groups, for instance, linking races or genders with negative features [65]. When the system identifies such biases, it corrects or retrains to provide children with equal and inclusive responses. Given that children will believe AI-generated responses to be authoritative, eliminating such biases at this point prevents the formation of skewed or disparate perspectives [27]. Furthermore, this AI-driven bias analysis serves to neutralize indirect misinformation, thereby guaranteeing long-term user trust and protecting a child's formative environment.

### 3.3. Cultural and Developmental Sensitivity

Fundamental elements of the framework are cultural awareness and age suitable of development. It guarantees that responses produced by artificial intelligence satisfy the particular cognitive and emotional needs of children at various age groups. While the framework stresses basic language and brief explanations for preschoolers (ages 6–12), it aims at nuanced, reflective responses that extend emotional intelligence and social awareness for teenagers (ages 13–17).When a small child asks about space travel, for example, the artificial intelligence has to react with a simple response grounded on logical analogues. One can provide an adolescent a long and more complex explanation. Here, cultural sensitivity also becomes a crucial factor since big language models have to incorporate many points of view into stories and stop feeding preconceptions that can offend specific groups. This feature adapts responses depending on the user's age and cultural background, therefore maintaining dialogues relevant and inclusive.

Simplifying dynamic material is changing language form and complexity to fit the developmental level of the user. Should a youngster indicate being eight years old, the algorithm reduces the reading level to early elementary school level, hence preventing "cognitive overload" [69]. Simplified writing not only increases involvement but also reduces the possibility of misinterpretation or false confidence. Not less important but older teenagers (13–17) can get more analytical, detail-rich answers that inspire critical thinking. Such calibration is essential since many children eventually lose interest or understand incomplete knowledge if they find advanced artificial intelligence outputs "too complex" [61].

Cultural sensitivity filters: Strict safeguards to eliminate or modify content that may be culturally offensive, trivializing, or exclusionary. For instance, if an LLM inadvertently uses stereotypes about a certain ethnic group, the filter intercepts it and re-contextualizes the response. Children from minority cultures have reported feeling alienated when AI chatbots deliver content that erases or misrepresents their backgrounds [74]. Adding cultural sensitivity ensures representation in stories, that is describing holiday traditions or historical figures, without causing harm or confusion.

Real-time parental monitoring dashboards: Provide parents with real time snapshots of AI interactions, flagging suspicious or damaging content for review. Parents are shocked that they cannot distinguish if their child is being instructed how to circumvent parental filters, viewing mature topics, or receiving hate messages [64]. Real-time monitoring lets parents spot if an artificial intelligence conversation is veering into dangerous territory—that of drug addiction or suicide—and act fast. Since adults know right away how the artificial intelligence is interacting with their children, this openness also helps to lower anxiety.

Ethical compliance scoring tools: Automated processes provide LLMs with an "ethical compliance" score for ongoing adherence to the guidelines. Cases of children exposed to AIgenerated content that includes sexual terminology or racial slurs demonstrate the need for ongoing monitoring [25], [62]. Ongoing scoring identifies potential backsliding or new patterns of bias falling through. The system triggers a review or retraining session when the score dips below a specified threshold. This step is particularly relevant given that parents often lack the time or technical literacy to audit every AI response, reinforcing the importance of a built-in safety net [65]. Through monthly analytics, flagged content is studied and corrected, upholding a consistent standard of cultural and developmental sensitivity.

### 3.4. Implementation of the Framework

Implementing the framework involves integrating real-time moderation tech into LLMs to identify and correct inappropriate output during engagement. Such technologies must also adapt dynamically in accordance with flagged content and provide real-time corrections to facilitate safe engagement. Also, data enrichment training needs to be used, where datasets are constructed with diverse cultural viewpoints and content that is suitable to the developmental needs of children. For instance, datasets would be audited to eliminate toxic stereotypes as well as incorporate educational content that fits the needs of children.

Also, content review processes should have human moderators who vet flagged responses and perform context-specific editing. There should be escalation procedures for interactions at high risk, e.g., signs of emotional distress or mental health crisis. Deployment consists of the creation of parental monitoring dashboards enabling parents to monitor interactions, filter out content, and be alerted of risky behavior. For example, parents can view summaries of interactions with AI such that they are better aware of their child's experience on the platform (See Figure 4).

The literature addresses the great issues raised by bringing LLMs to online environments for youngsters. Parents' most important concerns are biassed, offensive language, improper content for different age groups, and lack of adequate parental control choices and openness in contacts with artificial intelligence. Current studies show that situations in which children are exposed to abusive language, invited to engage in dangerous activities, or misled by false information are particularly common. Moreover, without parents' direct educational guidance, children's inclination to assign human-like qualities to artificial intelligence and develop emotional connections with chatbots poses novel psychological and ethical issues that usually fall beyond the parents' control.

**Figure** 4: Implementation of the Protection Framework for Safe Child-LLM Interaction

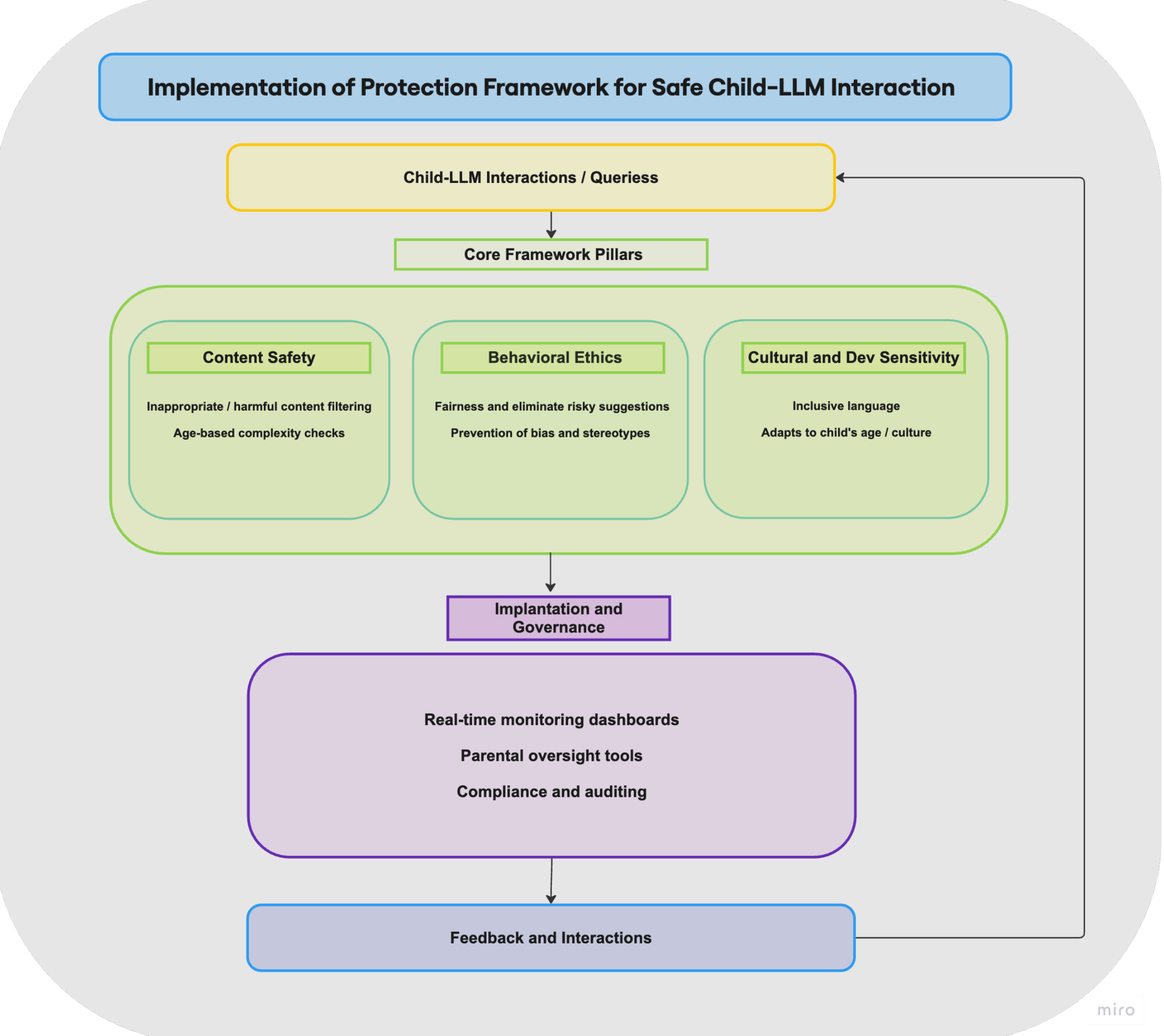

The suggested model anticipates comprehensive benefits in managing these challenges with an integrated system. It employs vigorous indicators of content safety to identify and eliminate improper or harmful content, thus maintaining ethical levels of conduct that permit the convergence of Ai generated content with familiar societal norms and values. Further, cultural and developmental sensitivity steps guarantee that our AI responses are customized to fit the varying requirements of children across various age ranges and cultures. Furthermore, our foundational framework places emphasize transparency and parental involvement, such as implementing through monitoring dashboards and real time alerts for risky behavior, which may make the difference between preventing kids from exposing personal information online to building trust for parents and peace of mind in responsible AI use. The necessity for change in the LLM landscape for kids' safety make the framework a significant step toward creating safer and more ethical AI systems for our next generation

# IV. Implications for Developers and Stakeholders

The model gives practical suggestions to LLM developers, the most important of which is incorporating robust content moderation capabilities, ethical compliance rating systems, and realtime monitoring mechanisms. With such functions, developers can make child's LLM use safe, trustworthy, and equitable. Parents, educators, and policymakers share an equally significant responsibility to provide a secure AI environment for children. Parents should have easy access to features like tracking dashboards and customizing content filters, which allow parents to keep an eye on the interactions of their children with AI. Teachers can integrate LLMs into the curriculum under strict adherence to the guidelines laid out in this framework so that these tools augment learning outcomes without undermining ethical and safety issues. They can also urge policymakers to adhere to the framework as well as advocate for policymakers to create child-specific data sets and develop regulatory jurisdiction over generative AI platforms that are designed to specifically target younger individuals. For developers, the framework provides a clear direction to produce safer and more ethical LLMs for children. Essential will be implementation of real-time content moderation technologies, dynamic simplification of information to enable age-friendly engagement, and automated bias testing technologies. These actions can reduce the chances of biassed or poisonous outcomes rather successfully. Creating varied training sets for cultural diversity and the avoidance of stereotypes must also be priorities for developers.

The safe implementation of artificial intelligence also depends much on non-developer stakeholders such as parents, educators, and policy-makers. Adaptable filters and interaction logs are basic tools parents need to monitor and limit their children's access to artificial intelligence. Teachers can utilize LLMs to aid in education but should be careful to ensure that there is adherence to guidelines in the framework to prevent dependence or dishonesty. Policymakers can inspire cooperation among developers, child welfare professionals, and instructors thereby ensuring adherence to such frameworks and supporting their expansion. Working together, these parties can guarantee a system whereby generative artificial intelligence supports rather than poses threats to young people.

### 4.1. Challenges and Limitations

Implementing the proposed framework is not trouble-free. It remains a significant technical challenge to detect malicious content in real time, especially context-dependent or ambiguous terms. LLMs must also be sensitive to diverse linguistic and cultural contexts, which implies enormous training data and ongoing updating to prevent biases. Compliance across cultures in international markets is very difficult because of varying cultural norms and legal standards. Further, a key limitation arises from the use of human moderators to review reported content that leaves room for the possibility of delays or inconsistencies, especially where time sensitivity is involved. Lastly, incorporation of dynamic content modification and ethics compliance software into current platforms may require significant development time and investment, which may be an impediment for widespread adoption.

The deployment of the proposed framework is faced with several challenges. A key challenge concerns the sophisticated technical conditions for real-time detection of harmful content. The token predictive nature of generative AI is prone to misinterpreting user input, leading to the generation of potentially harmful output, particularly in situations with contextual ambiguity. Second, the implementation of the framework to diverse cultural and linguistic contexts is a resource-demanding task that requires curating localized datasets in addition to addressing cultural nuances. Inclusivity with sensitivity to regional variation is an arduous, resource-demanding process.

One drawback is depending on human moderators to review flagged material, which can slow down the reaction to high-risk contacts. Another constraint of the system is scalability since using ethical compliance tools and dynamic content adaptation tools could call for significant technological and financial resources. Researchers should also investigate automated content moderation and bias detection tools to enable the system to be more scalable and efficient. Finally, ongoing discussion regarding ethical use of AI in child-centered settings must engage policymakers, developers, and educators so the structure is adaptable enough to include technology and societal innovation.

### 4.2. Future Directions

Subsequent studies should focus on refining the metrics of the framework through empirical evaluation and feedback cycles. Performing longitudinal studies to measure the effectiveness of the framework in various contexts would be informative on its practical implications. Other research directions involve developing improved datasets that more accurately capture the cognitive, emotional, and cultural requirements of children so that large language models' outputs can be inclusive and suitable for stages of development. Advancement in AI-powered content moderation and bias detection tools can also enhance the scalability and accuracy of the framework. Interdisciplinary collaboration among AI engineers, child psychologists, educators, and policymakers will be required to update and refine the framework periodically so that it remains relevant as technology evolves. Finally, there is a necessity to address ethical considerations of emotional attachment to artificial intelligence and implement protective measures for this eventuality to avoid long-term development and societal issues.

A further area for improvement is the creation of more representative datasets that better address the cognitive, emotional, and cultural needs of children. Collaboration with child development experts and cultural anthropologists can ensure the datasets are age-specific and encompassing. Researchers must also explore the application of automated content moderation and bias detection technologies to enable the framework to be more efficient and scalable. Finally, current controversies around using AI in an ethical way in child-focused settings must include policymakers, developers, and educators so that the framework is sufficiently flexible for technological advancement and societal needs.

## V. Conclusion

The analysis has identified substantial risks with children's use of LLMs found on parental concerns as well as empirical evidence. Parental concerns range from worry about exposure to inappropriate, biased, or culturally insensitive content to having no tools whatsoever to monitor or regulate such exposure. Empirical data supports these hazards; it shows that LLMs can generate poisonous output, magnify prejudice, and misinterpret information in ways that result in dangerous advise or destructive conduct. Along with the lack of strong content monitoring and data management openness, the vulnerability of children to trust LLMs and anthropomorphizing LLMs also generates these hazards. The results highlight the vital need of strong safety measures and systems meant to prevent children from these hazards as well as of encouraging responsible use of LLM technologies.

This study makes two primary contributions to the field. It initially conducts an extensive review of the safety and ethical concerns of children's use of large language models, synthesizing parent concerns and empirical research into a cohesive narrative. This therefore completes an important lacuna that has hitherto existed in the field of appreciating the various risks of LLMs. Second, the safety and ethics framework suggested above brings a novel method to measuring and enhancing the safety of child users of LLMs. By tackling content safety, ethical behavior, and cultural and developmental sensitivity, the framework provides concrete tools and guidance for developers, educators, and policymakers.

The evidence and framework presented in this document amount to a powerful demand for prioritization and swift action on the part of researchers, developers, and policymakers to place child safety at the top of the agenda when designing, developing, and regulating artificial intelligence systems. Large language models need to have mechanisms embedded to prevent harm, optimize positive learning experiences, and protect children's privacy and well-being. By promoting inter-disciplinary collaboration, the artificial intelligence research community can develop systems that amuse and inform while upholding the highest standards of safety and morality for the future generations of learners. The process is a foundation in addition to being a source of inspiration for constructive action so that future AI systems will educate children while upholding the greatest interests of the children and their parents.

**Conflict of interest**: The authors declare no competing interests.

**Acknowledgement**: This research is funded by NSF grants 2125858, 2236305, and the UT-Good Systems Grand Challenge. We would like to express our gratitude for these institutions' support, which made this study possible. Furthermore, in accordance with MLA (Modern Language Association) guidelines, we acknowledge the use of AI-powered tools, such as Anthropic's applications, for their assistance and support including in writing, editing, and improving the clarity and organization of the manuscript.